\documentstyle[12pt]{article}

\newcommand{\nc}{\newcommand}
\nc{\beq}{\begin{equation}}
\nc{\eeq}{\end{equation}}
\nc{\beqa}{\begin{eqnarray}}
\nc{\eeqa}{\end{eqnarray}}
\def\DS {D\!\!\!\!/}

\def\DM {\DS_{\, (\mu)}}

\input epsf
\newwrite\ffile\global\newcount\figno \global\figno=1

\def\writedef#1{}
\def\figin{\epsfcheck\figin}\def\figins{\epsfcheck\figins}
\def\epsfcheck{\ifx\epsfbox\UnDeFiNeD
\message{(NO epsf.tex, FIGURES WILL BE IGNORED)}
\gdef\figin##1{\vskip2in}\gdef\figins##1{\hskip.5in}% blank space instead
\else\message{(FIGURES WILL BE INCLUDED)}%
\gdef\figin##1{##1}\gdef\figins##1{##1}\fi}
\def\figinsert{}
\def\ifig#1#2#3{\xdef#1{fig.~\the\figno}
\writedef{#1\leftbracket fig.\noexpand~\the\figno}%
\figinsert\figin{\centerline{#3}}\medskip\centerline{\vbox{\baselineskip12pt
\advance\hsize by -1truein\center\footnotesize{  Fig.~\the\figno.} #2}}
\bigskip\endinsert\global\advance\figno by1}
\def\endinsert{}
\begin{document}

\title{\large{\bf On the QCD Phase Transition at
Finite Baryon Density }}

\author{
Stephen D.H.~Hsu\thanks{hsu@duende.uoregon.edu}
 \\ Department of Physics, \\
University of Oregon, Eugene OR 97403-5203 \\ \\
Myckola Schwetz\thanks{myckola@baobab.rutgers.edu}
 \\ Department of Physics and Astronomy, \\
Rutgers University, 
Piscataway NJ 08855-0849 \\ \\
}

\date{March, 1998}

\maketitle

\begin{picture}(0,0)(0,0)
\put(350,350){OITS-647}
\put(350,330){RU-98-10}
\end{picture}
\vspace{-24pt}

\begin{abstract}
We investigate the QCD chiral phase transition at 
finite chemical potential $\mu$, using the
renormalization group (RG) to characterize the
infrared behavior of sigma models constrained
by the flavor and spacetime symmetries. The
results are similar to those obtained from RG
analysis of the finite temperature transition at
zero baryon density. 
When there are more than two massless flavors 
of quarks, a first order transition is predicted for
the entire phase boundary. 
In the two flavor case, a boundary with first and
second order regions separated by a tricritical point
seems most likely.
We discuss the real-world case with two
light quarks and an intermediate mass strange quark.
Improved lattice data on the temperature transition
will strongly constrain the possibilities for the 
phase boundary.

\end{abstract}

\newpage

\section{Introduction}

\begin{figure}[htb]
\center{
\leavevmode
\epsfbox{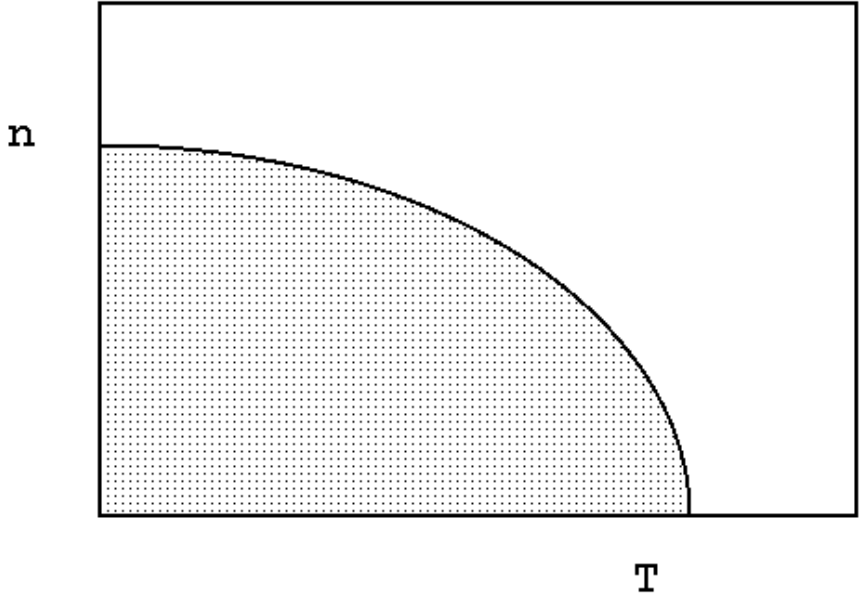}
\caption{Phase diagram of QCD for varying 
baryon density (n), temperature (T).
The shaded region exhibits chiral symmetry breaking.} 
\label{phasediag}
}
\end{figure}

Due to a technical obstruction, our theoretical understanding of 
QCD at finite baryon density remains limited. The basic difficulty is
that the introduction of a chemical potential leads to a complex  
effective action after integration over the quark fields. This creates
severe difficulties for lattice simulations \cite{lattice} and precludes
the use of rigorous inequalities \cite{ineq}. While intuitive arguments
suggest a phase diagram like that displayed in figure (\ref{phasediag}), 
with the quark condensate $\langle \bar{q} q \rangle$ playing the role of
the order parameter, in reality only the zero density axis of the diagram 
has been explored in any systematic way. In particular, little is known
about the properties of the phase transition at finite chemical potential
$\mu$.

This lack of theoretical results is particularly galling considering
the relevance of large quark density to topics such as cosmology,
astrophysics (neutron stars) and heavy ion collisions. Recent
work using an instanton-inspired model of quark interactions at
high density \cite{instanton} suggests some of the possible exotic
phenomena. 

In this letter we will apply the renormalization group (RG) and the
ideas of universality to the QCD chiral phase transition at finite $\mu$.
These methods have been successful in condensed matter physics \cite{condmat}
and have previously been applied to the finite temperature transition
in QCD \cite{finiteT}. The basic idea is to assume a second order transition
and attempt to find a sigma model description, consistent with the symmetries, 
of the relevant long-wavelength
degrees of freedom. In order that the description be self-consistent the
model must possess an infrared stable (IRS) fixed point. Lack of such a fixed
point signals an instability, and therefore a first order transition.
There are two logical weaknesses of this technique. One is that a 
non-perturbative
fixed point might exist, even when the perturbative beta function does not
exhibit one. In this case the transition would be second order, but in some
novel universality class which is difficult to analyze. Secondly, even when a
(perturbative) IRS fixed point exists in the sigma model, the dynamics of the 
underlying theory at shorter distances can still cause a first order 
transition to
occur before the basin of attraction of the fixed point is reached.  
A related point has been raised by Kocic and Kogut \cite{Kogut}, who point out
that the composite nature of the mesons, which is ignored in this analysis,
may affect the results.

\epsfysize=10.0 cm

Having stated these important caveats, we now proceed with the analysis.
Massless QCD at finite chemical potential is described by the partition
function
\beq
\label{partition}
Z = \int DA~ e^{-S[A]} ~ {\rm det} \DM~~,
\eeq
where 
\beq
\label{Dirac}
\DM ~=~ \DS + \mu \gamma_4 ~~.
\eeq
In our convention $\DS$ is hermitian and $\gamma_4$ antihermitian. The
eigenvalues $\lambda_i$ of $\DM$ are generally complex, leading to a 
complex quark determinant. Despite this, the partition function can
be rewritten as a sum over real (albeit possibly negative) terms. One
way to see this is to work in $A_0 = 0$ gauge, where direct examination
of the Dirac equation shows that the spectrum of eigenvalues 
$\{ \lambda_i \}$ associated with a gauge configuration 
$A_i (\vec{x}, x_4)$ is mapped to its complex conjugate 
$\{ \lambda_i^* \}$ when the gauge configuration is parity-inverted to
$ - A_i (- \vec{x}, x_4)$. 
(The original eigenspinors $\Psi_i ( \vec{x}, x_4 )$ are mapped to
$\gamma_2 \Psi^*_i ( - \vec{x}, x_4 )$ .)  
Since the original gauge field and its parity
partner have the same action $S[A]$, we can rewrite (\ref{partition})
as a real integral over pairs of gauge configurations.
The boundary conditions on the fields in the
time direction (periodic for bosons and anti-periodic for fermions)  
determine the temperature of the system. 

We are interested
in the effective, long-wavelength description of the theory as we approach
the phase boundary in the $( \mu , T )$ plane. Let us discuss the symmetry
constraints on this effective description. If we approach from the
direction of the phase with chiral symmetry breaking, the effective theory
must exhibit spontaneous breaking of $SU(N) \times SU(N)$ symmetry (here
$N$ is the number of massless quark flavors) with the corresponding
Goldstone degrees of freedom. Actually, the issue is somewhat more complicated
than this because of the anomalous $U_A (1)$ symmetry, which we discuss in the
next section.
As far as spacetime symmetries, the chemical potential
term breaks the Euclidean $O(4)$ symmetry down to $O(3)$. This allows a
more general form for the `time' derivative parts of the kinetic
energy term. In other words 
$ {\rm tr} ( \partial_\mu \Phi ) ^{\dagger} ( \partial_\mu \Phi )$ becomes 
\beq
\label{ab}
{\rm tr} ( \partial_i \Phi )^{\dagger}( \partial_i \Phi ) ~+~ 
a ( \mu, T )~ {\rm tr} (\partial_4 \Phi)^{\dagger} (\partial_4 \Phi)  
 ~+~ b_1 (\mu, T)~  {\rm tr}  \Phi^{\dagger} \partial_4 \Phi ~+~  
b_2 (\mu,T)~ {\rm tr} \Phi \partial_4 \Phi^{\dagger} 
,
\eeq
where $a (\mu = 0, T) = 1$ and $b_i (\mu = 0, T) = 0$. Physical arguments
suggest that $a (\mu, T)$ remains positive. Otherwise, $x_4$-dependent 
fluctuations of $\Phi$ would grow without bound. It is fortuitous,
and somewhat surprising, that the unknown coefficients in (\ref{ab}) 
are largely irrelevant to the RG analysis, as we
discuss below. 

At finite temperature the boundary conditions lead to a discretization
of the $x_4$ derivatives. Usually one assumes that only the static
mode is relevant to the RG analysis, leading to an effectively three
dimensional problem. However, this assumption needs to be reconsidered
here, if only because we might be interested in the transition at
zero temperature and finite $\mu$. When the temperature is exactly
zero the RG analysis must be performed in $d=4$ dimensions. 
It is clear that the introduction of the $a,b_i$ parameters will
affect the manner in which the non-static modes decouple from the
IR analysis. This can alter the dynamics of the model as the IR region
is apporached. However,
for any non-zero temperature (as long as $a,b_i$ are not precisely zero), 
one eventually reaches a scale at which the non-static modes are suppressed
and the dimensionality is effectively $d=3$. Unless $T = 0$, the 
test of the self-consistency of a second order transition remains the same.
We will return to the role of the $a,b_i$ coefficients in section 3.

\section{Anomalous $U_A (1)$ and eta meson}

To determine the precise flavor symmetry at the phase transition
we must understand the fate of the $U_A (1)$ axial symmetry.
Towards this goal, we consider the two-point functions for 
particles which are in the same $U(N) \times U(N)$ multiplet 
but {\it not} in the same $SU(N) \times SU(N)$ multiplet.
For $N > 2$ it is possible to show that the respective correlation
functions become identical as the chiral symmetry is restored.
This was originally demonstrated in \cite{Ua} (see also
\cite{Cohen}) with the high temperature phase in mind, but the 
proof applies equally 
to high densities. The only difference in the expressions
is that the Dirac eigenvalues become complex quantities.  
The result implies a degeneracy within the entire $U(N) \times U(N)$ 
multiplet, and a corresponding restoration of the $U_A (1)$
symmetry at the phase transition.

The two-point correlation functions for the $\pi$ and $\eta'$ are
given by
\beqa
\label{tp}
\langle \eta'(x) \eta'(0) \rangle ~&=&~
%\frac{1}{N_f}~
\langle ~\bar{\psi}_i \gamma_5 \psi_i~ (x)
{}~ \bar{\psi}_j \gamma_5 \psi_j ~ (0) \rangle \\
\langle \pi^a (x) \pi^a (0) \rangle ~&=&~
\langle~ \bar{\psi} \tau^a \gamma_5 \psi~(x)
{}~ \bar{\psi} \tau^a \gamma_5 \psi ~(0) ~\rangle
\eeqa
We can express these correlators in terms of
exact quark propagators 
\beq
S_A (x,y) ~=~ \sum_k { { \Psi^{\dagger}_k (x) \Psi_k (y)} \over 
\lambda_k - i m_q }~,
\eeq
where $A$ denotes the background gauge field in which the 
eigenvalues and eigenfunctions are computed.
One finds two types of contributions:
a disconnected contribution
\beq
\label{dc}
\frac{1}{Z}\int DA ~e^{- S[A]}~{\rm det} \DM~
{\rm tr}[\Gamma S_A(x,x)] {\rm tr} 
[\Gamma S_A (0,0)]
\eeq
and a connected part
\beq
\label{c}
\frac{1}{Z} \int DA ~e^{- S[A]} ~{\rm det} \DM 
~ {\rm tr}[S_A (x,0) \Gamma S_A(x,0) \Gamma].
\eeq
Here $\Gamma = \gamma_5$ for the $\eta'$ and
$\Gamma = \tau^a \gamma_5$ for the $\pi^a$.
The connected parts (\ref{c}) are identical since
$[ \tau^a, S_A ] = 0$.
For the pion, the disconnected part is zero since
${\rm tr} [ \tau^a] = 0$. Any $\eta'$-$\pi^a$ splitting is the result of
(\ref{dc}) for the $\eta'$.

Further analysis involves the careful consideration of contributions to
(\ref{dc}) from different sectors of the gauge field configuration
space with topological charge $\nu$. Only the non-zero $\nu$ sectors
can contribute to a splitting between the $\eta'$ and the pions. 
Working in the chirally restored phase, it can be shown that the 
contributions to the partition function 
from the sectors with non-trivial topology vanish 
like $m_q^{| \nu | N}$ as the quark mass 
$m_q$ approaches zero. (This is essentially a consequence of
the index theorem.) The zero-mode part of the two quark propagators
in (\ref{dc}) absorbs exactly two powers of $m_q$, which implies
that for $N > 2$ this contribution vanishes entirely when the 
quarks are exactly massless\footnote{Note
that this analysis applies only in the phase {\it without} chiral
symmetry breaking. In the broken phase the limit $m_q \rightarrow 0$
is more subtle \cite{Ua}.}.  
This result leads to the following conclusions:

\begin{itemize}

\item  $N = 2$:
The  $SU(2) \times SU(2)$ global symmetry is restored in
the high density/temperature phase. The $\eta$--$\pi^a$  
splitting is non-zero, 
but decreases smoothly to zero with temperature and density
as asymptotic freedom suppresses topological fluctuations. It remains
an open dynamical question whether the $\eta$ plays a 
role in the phase transition, and the relevant sigma model
has either a $U(2) \times U(2)$ or $O(4)$ flavor symmetry.

\item $N > 2$ : 
The $U(N) \times U(N)$ global symmetry is effectively restored
(up to high-dimension operators which are probably irrelevant
in the IR limit) in the high density/temperature phase. 
If the transition is continuous, the $\eta'$
becomes degenerate with the $\pi^a$'s at the phase boundary. 
The effective models of this chiral phase transition must
incorporate a $U(N) \times U(N)$ global symmetry.
(Note that the large-$N_c$ limit with any number of flavors
falls into this class.)

\end{itemize}

\section{Sigma models and RG flows}

Having identified the relevant symmetries constraining our sigma
models we can now proceed with the analysis of RG evolution in the
infrared. 
The critical behavior of the $U(N) \times U(N)$
linear sigma model has been studied in
$4 - \epsilon$ dimensions \cite{finiteT,ps}.
The most general renormalizable potential consistent
with the symmetries is 
\beq
U( \Phi ) ~=~ \frac{1}{2} m_{\Phi}^2 ~ {\rm tr}~ \Phi^{\dagger} \Phi
~+~ g_1 ( {\rm tr}~ \Phi^{\dagger} \Phi)^2 ~+~ 
g_2~ {\rm tr}~ ( \Phi^{\dagger} \Phi )^2 ~~~.
\eeq
The one loop $\beta$-functions for $g_1$ and $g_2$ are
\beqa
\label{beta}
\beta_1 ~=~ - \epsilon g_1 ~+~ {N^2 + 4 \over 3}g_1^2 ~+~ 
{4 N \over 3} g_1 g_2 ~+~ g_2^2~,\\
\beta_2 ~=~ - \epsilon g_2 ~+~ 2g_1 g_2 ~+~ {2 N \over 3} g_2^2 ~~.
\nonumber ~~~~~~~~~~~~~~~
\eeqa
The stability of a fixed point $g^*$ (zero of (\ref{beta})) is 
determined by the presence of real and positive eigenvalues for 
the matrix $w_{ij}= \partial \beta_i / \partial g_j$ at $g=g^*$. 
The corresponding analysis has been done before for finite
temperature ($T \neq 0$) and zero baryon density ($\mu=0$), 
where the effective dimensionality is $d=3$
($\epsilon = 1$). 

For $N > \sqrt{3}$ there is no infrared-stable fixed point
with $g_1, g_2 \sim O(\epsilon)$. For example,
when $g_2 = 0$ the system effectively
becomes the $O(2N)$ linear sigma model. But when both couplings
are present the fixed point with $g_2^* = 0$ is
unstable in $g_2$ direction. Therefore the phase transition is predicted
to be of the first order.

The case with $N = 2$ is more complicated because of the status of
the eta meson. If the eta meson becomes massless at the transition,
there is a $U(2) \times U(2)$ symmetry. The RG equations are those
of (\ref{beta}), and for $N=2$ they have no IRS fixed point.
Otherwise, the relevant model is the $O(4)$ sigma model 
(n=4 isotropic Heisenberg magnet) with only one coupling. The RG analysis 
exhibits an IRS fixed point and the possibility of a second order 
transition.

Now, consider non-zero $\mu$.
If the temperature is also non-zero, the effective dimensionality 
is $d=3$,  and the $a,b_i$ terms in (\ref{ab}) play no role, 
as they only affect non-static modes.
The analysis in this case is therefore identical to that already
performed. In other words, the universality
classes available to describe the phase transition along the entire
boundary in figure (\ref{phasediag}), except near zero temperature,
are precisely the same as for the transition on the zero density
line.

The case of zero temperature is distinct, as there is no discretization
of the $x_4$ modes. We must therefore retain the $a,b_i$ terms and examine
their effect on the $d=4$ beta functions.  Simple calculation (at one loop)
shows that the new beta functions will be identical, up to an overall
rescaling, to those of (\ref{beta}) with $\epsilon$ set to 0. 
The one loop analysis
is straightforward, requiring only the evaluation of the ``fish'' diagram.
The logarithmic divergence in that graph is changed by an overall constant
proportional to $~{1 \over \sqrt{a}} \arctan ( \frac{1}{\sqrt{a}} )~$, 
but independent of $b_i$. For $N > 2$ there is still no IRS fixed point,
and the transition is therefore predicted to be first order on the entire
boundary.  There is also the
possiblity of massless fermionic modes which are relevant to the
transition.
The possibility of massless fermions here is unlikely, as
one cannot satisfy the 'tHooft anomaly matching conditions with massless
color singlet
fermions when $N > 2$ \cite{TH}. In any case, the addition of massless baryons 
coupled to $\Phi$ does not stabilize any fixed point at the one loop level.

For $N=2$ there is an IRS Gaussian fixed 
point which could model the second order transition.
This leaves open the possibility that the entire phase
boundary is second order. For this to be the case
there would be a smooth interpolation from the $\epsilon = 1$ to
the $\epsilon = 0$ critical behavior as the temperature approaches
zero. This seems implausible to us, as it would require a new family
of universality classes with critical behavior intermediate between
the Gaussian and $O(4)$ fixed points. However, we have no solid
evidence to rule out this possibility.
In the two flavor case the anomaly matching conditions allow
massless fermionic degrees of freedom (parity-doubled baryons) 
which are relevant to the transition. Including these degrees of freedom
leads to a Higgs-Yukawa model like that of the standard model
with zero gauge couplings. Triviality of this system again implies
a Gaussian fixed point in the IR \cite{Montvay}.

\section{Discussion}

It is somewhat surprising that the introduction of
a chemical potential has little effect on the types of sigma
models that could govern the QCD phase transition.
We should clarify that this is far from a statement that finite
baryon density has no effect on the phase transition. 
In most cases the transition is predicted to be 
first order, with  characteristics such as latent
heat and size of discontinuity in order parameter which are
presumably
strongly dependent on $\mu$. In the case of two flavors, however,
there is the intriguing (though implausible) 
possibility that the transition remains
second order at high baryon density. Current lattice data is 
consistent with a $\mu = 0$, finite T transition which is second
order and in the universality class of the O(4) sigma model
\cite{exp}. This critical behavior could persist along 
the entire boundary in figure  (\ref{phasediag}), 
except at the $T=0$ endpoint, where the fixed 
point becomes Gaussian. 

More likely in the two flavor case is that the transition
switches from second to first order at some point along the 
$(\mu,T)$ boundary. This could occur if, for example, the eta
meson becomes light enough to change the behavior from
$O(4)$ to $U(2) \times U(2)$. A better understanding of the
behavior of the topological susceptibility, and hence the eta
mass, at finite baryon density might help to decide this issue.
Recent model calculations using instanton-induced interactions
\cite{BR} and random matrix techniques \cite{SUNY} 
suggest that the $T=0$ transition is first order.
A transition between first and second order
behavior along the boundary in figure 1 would imply a
tricritical point. The inclusion of non-zero light quark masses 
would presumably smooth the second order transition to a
crossover, while the first order boundary would remain qualitatively
the same. The tricritical point would then become a critical point
(second order phase transition)
at which the line of first order transitions terminates.
Near this critical point the system would exhibit large fluctuations.

We now turn to a discussion of real-world QCD. An important
input into this discussion is the character of the real-world
finite temperature transition. Unfortunately, there is
disagreement on this issue between lattice groups using
Kogut-Susskind fermions (Columbia) \cite{Lattice1} and Wilson fermions 
(JLQCD) \cite{Lattice2}, 
with the
former predicting a smooth crossover and the latter a first order
transition. In the most recent simulations of the two flavor case 
the Wilson method is seen to reproduce the $O(4)$ 
critical exponents,
while the Kogut-Susskind method does not \cite{Lattice3}.
In what follows we will discuss the implications of both possibilities.

Consider how the nature of the
phase boundary changes as we increase the strange quark mass from zero
to infinity, as shown in figure 2. In this diagram the vertical axis
is strange quark mass and the horizontal axis is the phase boundary
itself as a function of baryon density $n_B$ (ie the projection of the
boundary from figure 1 onto the $n_B$ axis).
At $m_s = 0$ we have a three flavor model and the boundary
is predicted to be entirely first order. However, as we increase $m_s$
the small-$n_B$ part of the boundary must disappear, replaced by a smooth 
crossover.  For this to happen, at some intermediate
value of $m_s$ a critical point must appear on the boundary, separating
the first order and crossover behaviors. This critical point presumably
first appears at zero $n_B$ and migrates to larger $n_B$ as $m_s$ is
increased. If the massless two flavor boundary has a tricritical point,
the line of critical points (heavy line) in figure 2 will terminate at the point $A$.
Otherwise the endpoint is at some $A'$ on the far right of the diagram,
signalling the absence of any phase boundary above some critical $m_s$. 
The position of the real-world value of $m_s$ on this diagram is currently
unknown, pending better lattice simulations. If the Columbia group is correct,
and the zero density transition is a crossover, then the value of $m_s$ is as
drawn in figure 2 and the real-world phase boundary is likely to have a
critical point. If JLQCD is correct, and the zero density transition is first
order, the $m_s = 150$ MeV line is much lower on the diagram and the entire
phase boundary is first order.

\epsfysize=12.0 cm
\begin{figure}[htb]
\center{
\leavevmode
\epsfbox{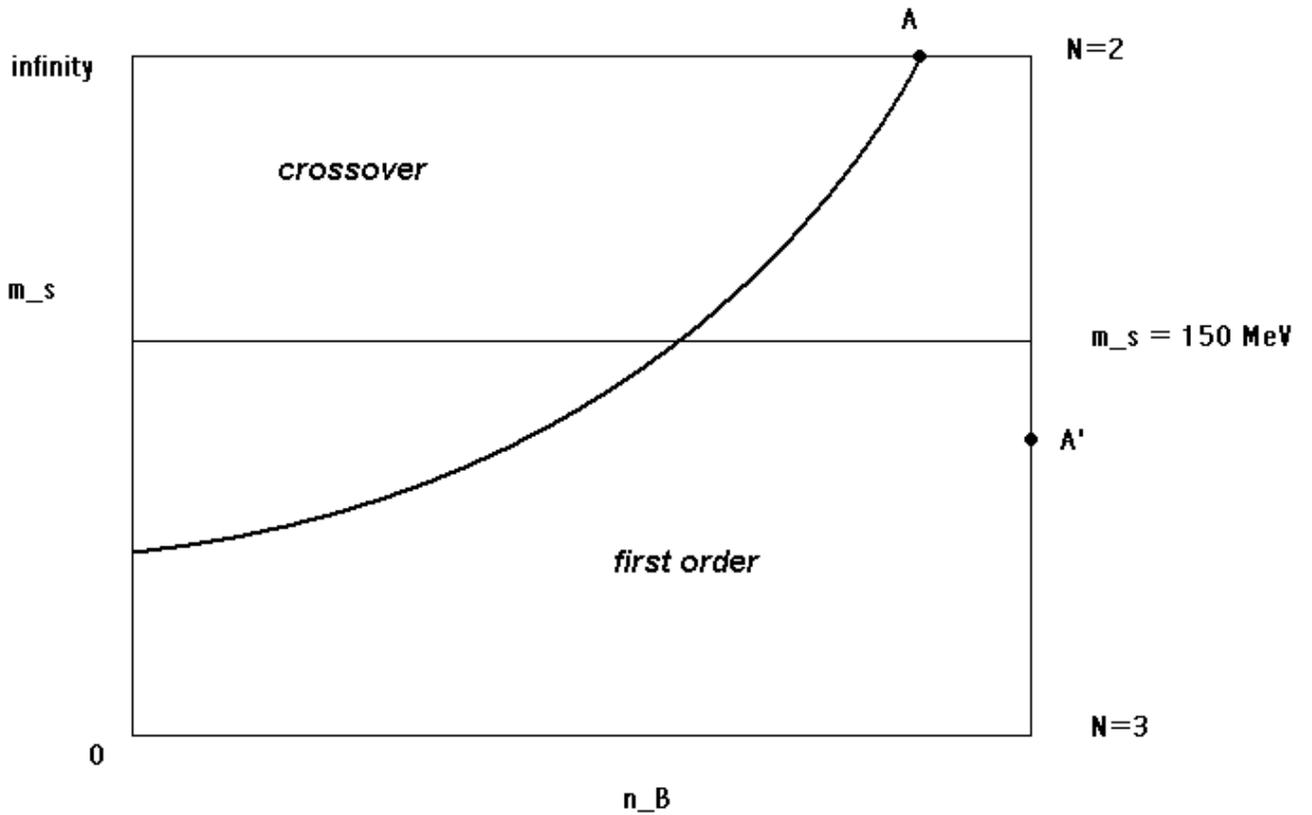}
\caption{Nature of QCD phase boundary for varying strange quark mass ($m_s$)
and baryon density ($n_B$). The heavy line terminating at A is a line
of second order transitions (critical points).} 
\label{phasediag2}
}
\end{figure}

\bigskip
\noindent 
The authors would like to thank Tetsuo Hatsuda, Rudy Hwa and Larry Yaffe
for useful discussions and comments. SDH is particularly grateful to
Tetsuo Hatsuda for helping to clarify the current status of lattice
simulations. This work was supported in part under DOE contracts 
DE-FG02-91ER40676 and DE-FG06-85ER40224.

%\newpage

\end{document}